\documentclass[%
 reprint,
 superscriptaddress,
 twocolumn, 
 amsmath,amssymb,
 aps,
 prb,
]{revtex4-2}

\usepackage[
 colorlinks = true,
 linkcolor = blue,
 urlcolor = red,
 citecolor = red,
 anchorcolor = blue
 filecolor=red]{hyperref}
\usepackage[T1]{fontenc}
\usepackage[utf8]{inputenc}
\usepackage{graphicx}
\usepackage[normalem]{ulem}
\usepackage{dcolumn}
\usepackage{color}
\usepackage{amsmath}
\usepackage{braket}
\usepackage{amssymb}
\usepackage{appendix}
\usepackage{MnSymbol}
\usepackage{orcidlink}
\usepackage{glossaries}
\newacronym{sig}{$\sigma_{1}(\omega)$}{the optical conductivity}
\newacronym{MIR}{MIR}{mid-infrared}
\newacronym{FL}{FL}{Fermi liquid}
\newacronym{MFL}{MFL}{marginal Fermi liquid}

\glsunset{sig}

\begin{document}

\title{Optical response of quantum metals}

\author{Erik van Heumen \orcidlink{0000-0001-9536-578X}}
\affiliation{Institute of Physics, University of Amsterdam, Science park 904, 1098 XH Amsterdam, The Netherlands}

\author{Martin Dressel \orcidlink{0000-0003-1907-052X}}
\affiliation{1. Physikalisches Institut, Universität Stuttgart, Pfaffenwaldring 57, 70569 Stuttgart, Germany}

\author{Simone Fratini \orcidlink{0000-0002-4750-3241}}
\affiliation{Universit\'e Grenoble Alpes, CNRS, Grenoble INP, Institut N\'eel, 38000 Grenoble, France}

\date{\today}

\begin{abstract}
Strongly interacting quantum systems display puzzling electrical transport properties whose interpretation remains controversial \cite{Cooper09,Basov:RMP2011,Bruin,analytis2014,Brown,cao2020,nguyen2021,PhillipsScience22,lee2023,shekhter2026,xia2026,Priya2027}. 
Many theoretical and phenomenological analyses assume that the complex behaviour observed in such `quantum materials' originates either from quasi-particles dressed with an anomalous electron self-energy \cite{DMFTReview,MichonNatComm23}
or from collective excitations associated with hydrodynamic flow \cite{baggioli2023,hui2025}. Despite their distinctly different starting points, the underlying assumption is that dissipation of electrical current arises from the relaxation of the total momentum of charge carriers. This classical description breaks down in the absence of long-lived charge carriers \cite{VarmaMFL, SenthilPRL03,cubrovic2009,Else:PRL2021,PatelScience23,li2024}. In such strange metals dissipation seems to be governed by the rate at which equilibrium is thermally restored, and it has been suggested that this constitutes an upper bound, known as the `Planckian limit' \cite{Zaanen04,Hartnoll,Hartnoll_coll,FratiniPRL26,FratiniNPhys26}.

Here we show that in this quantum regime a description of the conductivity in terms of relaxation rates no longer suffices: strange metal transport is carried instead by fast-decaying current fluctuations, which together with quantum detailed balance give rise to the widely observed Planckian dissipation. The distinction between classical and quantum metals becomes transparent when experimental data is analysed in terms of the current-current correlation function instead of the optical conductivity. When applied to the cuprate superconductors, such analysis uncovers that the strange metal response is present at all doping levels, and that a second, classical component develops above optimal doping. This simple, but powerful, approach further highlights essential features associated with superconductivity, (pseudo-)gaps, provides fresh insight in the properties of metals, and identifies key areas where progress in our theoretical understanding is needed. 

\end{abstract}

\maketitle

The remarkable applicability of the Drude model to the electrical conductivity of conventional metals is a consequence of the classical nature of charge transport. In such metals, the resistivity or optical conductivity is governed by the relaxation of total momentum associated with the inelastic scattering of electrons \cite{GrunerDresselBook02}. Even in extreme cases, such as the non-degenerate or hydrodynamic limit, the optical conductivity will feature a characteristic Drude peak \cite{baggioli2023,hui2025}.

Deviations from a simple Drude response can already emerge in the regime of weak to moderate interactions and optical spectroscopy data have been used to extract crucial information, such as the electron-boson coupling strength or the energy scale of the interactions governing the Fermi liquid \cite{allen1971,maslov2012,berthod13}, with the potential to provide quantitative information on the interactions responsible for superconductivity \cite{farnworth1974,Carbotte1999}. 

The anomalous electrical properties of the cuprates have led to characterizations such as `bad', `incoherent' or `strange' metals, which has carried over to other materials such as organics \cite{Priya2027}, nickelates \cite{lee2023}, twisted bi-layer graphene \cite{cao2020}, dichalcogenides \cite{xia2026}, pnictides \cite{analytis2014}, optical lattices \cite{Brown} and others \cite{Bruin,nguyen2021,PhillipsScience22}. For the cuprates in particular, one finds that at high temperature the effective mean free path appears shorter than the interatomic distance in violation of the Ioffe-Regel limit \cite{emery1995}. Another surprise is that the resistivity remains linear to the lowest measured temperatures \cite{Cooper09}. 

It appears likely that in such strange metals the assumptions underlying conventional approaches to charge transport are no longer valid. In the cuprates, the absence of quasi-particles has long been suspected \cite{VarmaMFL, emery1995}, and reported by angle resolved photoemission spectroscopy \cite{damascelli2003angle-resolved}. More recent experiments point to a partial restoration of quasi-particles above a hole doping $p^{*}\approx 0.2$ \cite{fournier2010,Chen:science2019}, although the strange metal persists alongside the classical metal component \cite{Ayres21}. Such experiments have spurred the development of alternative theoretical approaches to describe electrical transport in the absence of quasi-particles in a variety of materials \cite{VarmaMFL,emery1995,cubrovic2009,PatelScience23,li2024}. A common feature that has emerged in a wide range of settings, is that the total momentum associated with electrical current seems to relax at the rate of thermalisation, $\hbar/\tau\sim k_{\text{B}}T$ \cite{Hartnoll,gouteraux2016,Zaanen18,PatelScience23,balm2023,li2024}.

Under these circumstances, the approximate connection between conductivity and relaxation time is unlikely to prevail. This is because in the quantum regime, dissipation (encoded in the finite electrical conductivity of a metal) differs from fluctuation (the relaxation and variation of current at the microscopic level) through a quantum detailed balance factor \cite{Mahan}. The lack of consensus among different experimental techniques is perhaps the best indication that the conventional framework does not apply to strange metals. For example, the magnitude of the `Planckian' dissipation rate inferred from transport experiments using the classical Drude formula \cite{Hartnoll_coll} is inconsistent with the width of the Drude-like peak observed in optical experiments \cite{vanHeumen22}, and neither of those agree with the single particle scattering rates extracted from photoemission spectra \cite{PhillipsScience22}.\\
\\
\begin{figure*}[t]
    \centering
    \includegraphics[width=18cm]{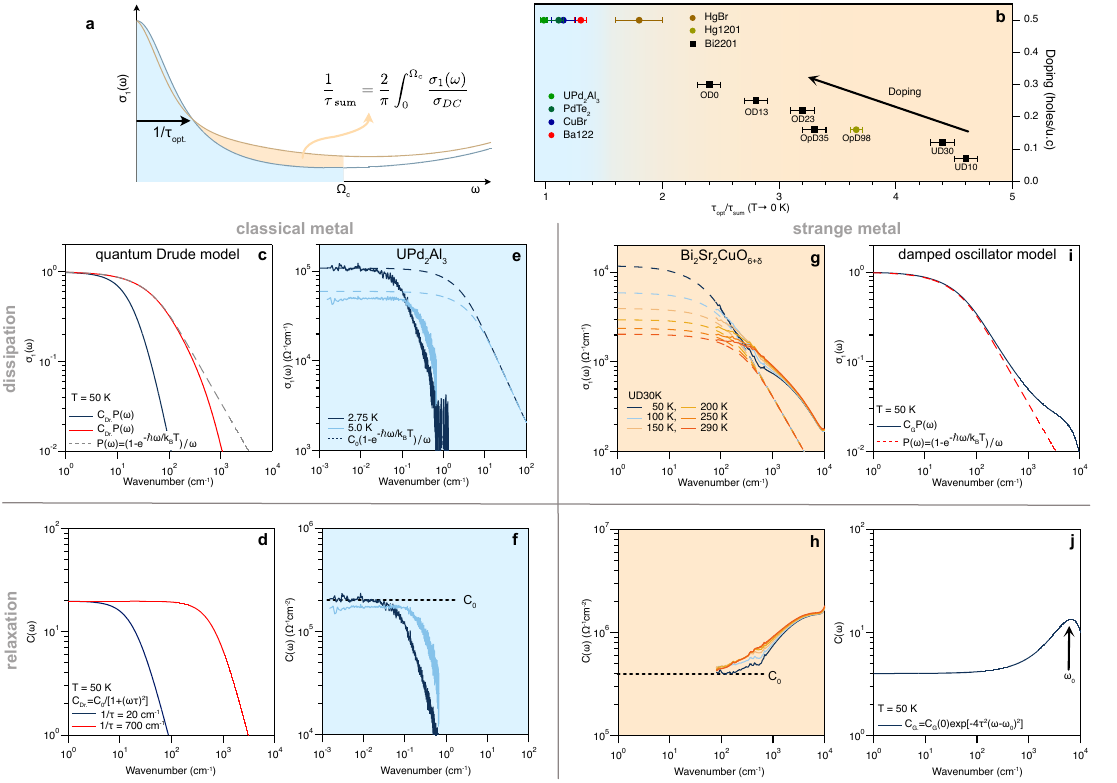}
    \caption{\textbf{Optical response and current correlations from classical to quantum metals.} 
    \textbf{a}: The optical conductivity, \gls*{sig}, of classical metals reaches a maximum at zero frequency with width $1/\tau_{\text{opt}}$ (blue curve/shade). This width coincides with the integral width $1/\tau_{\text{sum}}$, when there are well-defined current carrying excitations that decay at a rate slower than the Planckian bound. In strongly correlated metals a stretched \gls*{sig} is often observed and $1/\tau_{\text{sum}}$ exceeds $1/\tau_{\text{opt}}$ (yellow curve/shade).
    \textbf{b}: The ratio $\tau_{\text{opt.}}/\tau_{\text{sum}}$ for a range of different metals classifies strongly correlated metals as having $\tau_{\text{opt.}}/\tau_{\text{sum}}\gg 1$. The Bi$_{2}$Sr$_{2}$CuO$_{4+\delta}$ series with various dopings are arranged according to the right axis. Here OD, OpD and UD refer to overdoped, optimally doped and underdoped compounds with the respective $T_c$; see also Fig.~\ref{fig:Fig2} and Extended Data Fig.~\ref{figE:sigSM}. 
    \textbf{c}: For classical metals \gls*{sig} features a Drude-like conductivity (black) with $1/\tau<k_{\text{B}}T$. When $1/\tau$ increases (red line), \gls*{sig} approaches $P(\omega)$ (grey dashed line).
    \textbf{d}: The corresponding current-current correlation functions, $C(\omega)$, is a Lorentzian centred at zero frequency, reflecting the exponential decay of $C(t)$ in the time domain.
    \textbf{e}: \gls*{sig} of UPd$_2$Al$_3$ is the prime example of a Drude response \cite{Scheffler2005,Scheffler2010}. 
    \textbf{f}: The current correlations described by $C(\omega)$  are similar to \textbf{d}, indicating an exponential decay in the time domain. While the charge carriers in UPd$_2$Al$_3$ are strongly renormalized, it is a classical metal where dissipation is governed only by momentum relaxation.
    \textbf{g}: \gls*{sig} of Bi$_{2-x}$Pb$_{x}$Sr$_{2-y}$La$_{y}$CuO$_{4+\delta}$ (UD30) measured at indicated temperatures \cite{vanHeumen22}. $P(\omega,T)$ is obtained by taking $C_{0}$ (panel \textbf{h}) multiplied by the quantum detailed balance factor in Eq. \ref{eq:Kubo}. 
    \textbf{h}: $C(\omega)$ is dominated by a broad spectrum in the mid-infrared with a frequency-independent tail to zero frequency. The extrapolated value of this tail, indicated as $C_{0}$ determines the coefficient of Planckian dissipation. \textbf{i,j}: \gls*{sig} (\textbf{i}) for a $C(\omega)$ spectrum with a single Gaussian peak (\textbf{j}) qualitatively matches the strange metal conductivity presented in panel \textbf{g}.}
    \label{fig:Fig1}
\end{figure*}

\noindent\textbf{A measure of strangeness.}\\
In classical metals, current dissipates through the relaxation of total momentum. Irrespective of the precise mechanism the optical conductivity (and, by extension, the resistivity) will be dominated by a Drude-like peak \cite{Hartnoll,baggioli2023,vanHeumen22,FratiniPRL26}. Quite generally, one thus expects that the width, $1/\tau$, of the $\omega \approx 0$ absorption is related to the DC conductivity according to, 
\begin{equation}
    1/\tau_{\text{sum}}=\frac{\omega_{p}^{2}}{4\pi\sigma_{DC}}=\frac{2}{\pi}\int_{0}^{\Omega_{c}}\,{\rm d} \omega\,\frac{\sigma_{1}(\omega)}{\sigma_{DC}} \quad ,
\end{equation}
where the integral is the partial sum rule integrating over the free charge response (Fig. \ref{fig:Fig1}A). 
In metals where coherent, current carrying excitations exist, $1/\tau_{\text{sum}}$ should be close to the momentum relaxation rate that can be read directly from the width, $1/\tau_{\text{opt}}=1/\tau_{\text{opt}}$ of the experimentally measured $\omega \approx 0$ absorption peak. 
 
Figure \ref{fig:Fig1}\textbf{b} features several classical metals (contained in the blue shaded area), where modifications of the Drude response are small and well described within a memory function formalism \cite{gotze1972}. In these materials, $\tau_{\text{opt}}/\tau_{\text{sum}}\approx 1-2$, reflecting a moderate transfer of spectral weight to a `foot' in \gls*{sig} at higher energy \cite{berthod13} (see also Extended Data Fig.~\ref{figE:decay}). The situation is very different in strange metals where we find that $\tau_{\text{opt}}/\tau_{\text{sum}}\gg 1$ instead (highlighted by the yellow shaded area in Fig. \ref{fig:Fig1}\textbf{b}). 
As we proceed to show, in such cases it is particularly instructive to  express the experimental data  in terms of current-current correlation functions instead of the usual optical conductivity.\\

\noindent\textbf{Understanding optical absorption from current correlations}\\ 
In the following we will use the shorthand $C(t)\equiv\langle J(t)J(0)\rangle$ to denote the current-current correlation function and $C(\omega)\equiv{\rm Re}\int_0^\infty {\rm d}t\, e^{{\rm i}\omega t} C(t)$ for its Fourier transform. The Kubo formula connects the optical conductivity to $C(\omega)$ according to \cite{Mahan},
\begin{equation}
\sigma_1(\omega)= \frac{1-e^{-\hbar \omega/k_BT}}{V \hbar \omega}C(\omega) = P(\omega)\,C(\omega)\quad ,
    \label{eq:Kubo}
\end{equation}
with $V$ the system volume. This equation holds exactly for quantum systems in thermodynamic equilibrium. The frequency-dependent prefactor multiplying the correlation function is the quantum generalization of the $1/k_BT$ factor that appears in classical response functions: it entails the Kubo-Martin-Schwinger (KMS) detailed balance condition and the resulting fluctuation-dissipation theorem. A short derivation is provided in the supplementary material.
Seemingly, equation \eqref{eq:Kubo} is an innocuous transformation from \gls*{sig} to $C(\omega)$, but it has major consequences on the interpretation of experiments in quantum materials. It distinguishes dissipation from relaxation, and cleanly separates materials with long lived excitations from those without.

In the classical Drude treatment of transport as well as in its semi-classical Sommerfeld generalization one finds that the current correlations decay exponentially in time as $C(\mathrm{t})\simeq C_0 e^{-\mathrm{t}/\tau}$ as long as scattering events are uncorrelated, with $\tau$ the relaxation time \cite{GrunerDresselBook02}. Fourier transforming $C(t)$ yields a Lorenztian shape for $C(\omega)$ (Fig. \ref{fig:Fig1}\textbf{d}). Further taking the classical limit $\hbar\to 0$, where $P(\omega)=1/k_BT$ is a constant, leads to a Lorentzian shape for \gls*{sig} too. We can straightforwardly generalise the Drude-Sommerfeld model to the quantum realm by including the full frequency-dependence of $P(\omega)$ in equation \eqref{eq:Kubo}. The corresponding \gls*{sig} recovers the usual Drude peak as long as $\hbar/\tau\ll k_{\text{B}}T$  (Fig. \ref{fig:Fig1}\textbf{c}): this condition  defines classical metals \cite{Ashcroft}. 

When this condition is not fulfilled, the proposed quantum extension of the Drude model seamlessly connects to the notion of a Planckian bound on dissipation \cite{Zaanen04,Hartnoll}. Decreasing the lifetime of the carriers will initially broaden the optical conductivity peak as expected. However, as $\hbar/\tau\rightarrow k_{\text{B}}T$ 
the  prefactor $P(\omega)$ in equation \eqref{eq:Kubo} takes over, and making $\hbar/\tau$ larger (red curve, Fig. \ref{fig:Fig1}\textbf{c}) by increasing the rate of collisions will not broaden the conductivity any further, as conduction becomes dominated by quantum uncertainty \cite{FratiniPRL26,Ashcroft}. This limiting case is shown by the grey dashed line in Fig. \ref{fig:Fig1}\textbf{c}. Experimentally, quantum dissipation can be distinguished from the Drude result by an $\omega^{-1}$ decay rather than the usual $\omega^{-2}$ and it has a width (HWHM) $1/\tau_{\text{P}}\approx 1.59k_{\text{B}}T$, where the coefficient is close to those generally observed for strange metals \cite{Hartnoll_coll,Bruin}. This result is general: modifications of exponential decay can occur in various physical situations, as illustrated in Extended data figure \ref{figE:decay}. However, as long as the typical relaxation time is shorter than the thermalisation time, the width of \gls*{sig} will always be bounded by $\approx k_{\text{B}}T$.\\

\begin{figure*}[t]
    \centering
    \includegraphics[width=16cm]{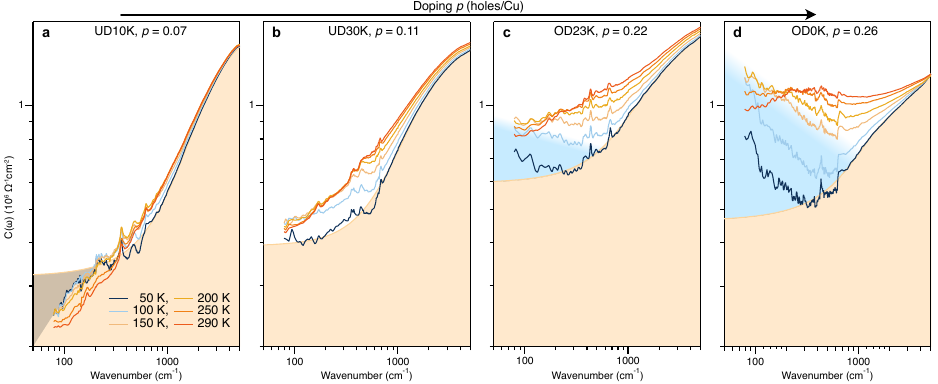}
    \caption{{\bf From strange to normal metal in cuprate superconductors.} Current-current correlation function extracted from the optical spectra of Bi2201, showing a smooth evolution from strange to normal carriers. From left to right,
    \textbf{a}: underdoped, T$_{c}$\,=\,10 K, 
    \textbf{b}: underdoped, T$_{c}$\,=\,30 K, 
    \textbf{c}: overdoped, T$_{c}$\,=\,23 K and 
    \textbf{d}: overdoped, T$_{c}$\,=\,0 K. The yellow shaded areas are obtained by fitting two Gaussian peaks to the $T =$ 50 K data between 200 cm$^{-1}$ and 7000 cm$^{-1}$. The darker shaded area in panel \textbf{a} indicates a suppression of spectral weight due to the possible opening of the pseudo-gap. The blue areas in panels \textbf{c} and \textbf{d} mark the classical carrier response.}
    \label{fig:Fig2}
\end{figure*}

\noindent\textbf{Classical vs. quantum metals}\\
To illustrate the utility of the current-current correlation function, we take as a first example the heavy fermion material UPd$_{2}$Al$_{3}$, where
hybridization of the localized $f$-states with the conduction electrons gives rise to strongly renormalized quasi-particles with an effective mass $66\,m_{\text{e}}$. The energy scale is reduced accordingly, which makes it possible to measure the real and imaginary parts of the conductivity independently of one another \cite{Dressel2006}.
As Fig.~\ref{fig:Fig1}\textbf{e,f} show, despite this strong mass renormalization the spectra are dominated by long-lived excitations to which a scattering rate $\hbar/\tau\ll k_BT$ can be assigned: both \gls*{sig} and $C(\omega)$ feature a $\omega\approx 0$ peak that can be successfully described using the Drude model \cite{Scheffler2005}.

We next consider an example where the transformation from \gls*{sig} to $C(\omega)$ provides new insight: the strange metal phase in cuprate superconductors. The optical conductivity for a close to optimally doped sample from the Bi$_{2}$Sr$_{2}$CuO$_{6=\delta}$ family (UD30, Fig. \ref{fig:Fig1}\textbf{g}) features an absorption peak centred at zero frequency \cite{vanHeumen22}. In this case, the peak at low frequency and low temperature is bounded by $P(\omega)$ (dashed lines) but significant spectral weight appears at higher energy. Transforming the data to $C(\omega)$ (Fig. \ref{fig:Fig1}\textbf{h}), we discover the key distinction with the classical case: in strange metals $C(\omega)$ is characterized by the absence of slowly relaxing excitations at $\omega\approx 0$. These are replaced by a broad, high energy spectrum that becomes nearly flat and temperature independent at low frequency, in particular in the relevant region $\hbar\omega\lesssim k_{\text{B}}T$. Taking the limiting value $C_{0}$ for $\omega\rightarrow 0$, equation \eqref{eq:Kubo} predicts quantitatively the temperature dependence of the low frequency optical conductivity (dashed lines in Fig. \ref{fig:Fig1}\textbf{g}) and the linear temperature dependence of the resistivity, as $\rho(T)=k_BT/C_0$ \cite{FratiniNPhys26,FratiniPRL26,Prelovsek-EPL01}. We relate this low frequency asymptote with the strange metal behaviour observed in transport.
A simple phenomenological description is obtained by assuming an overdamped harmonic relaxation $C_G(t)\sim e^{-t^2/\tau^2}\cos \omega_0 t$ (Fig. \ref{fig:Fig1}\textbf{i,j}), that qualitatively reproduces the broad maximum of $C(\omega)$ at a characteristic energy scale $\hbar\omega_{0}$, as well as the resulting shoulder observed in \gls*{sig} (Fig. \ref{fig:Fig1}\textbf{g}). It has been shown recently that such characteristic structure should emerge in doped Mott insulators at energy scales set by the magnetic exchange $J$ and the hole tunnelling $t$ \cite{FratiniNPhys26}, reflecting the composite structure of the charge carriers \cite{chalopin2026}.

Based on the foregoing discussion, we define quantum metals as all those cases where the quantum detailed balance prefactor in equation~(\ref{eq:Kubo}) qualitatively differentiates \gls*{sig} from $C(\omega)$, as in figures \ref{fig:Fig1}\textbf{g,h}. This will be the case when the narrow peak in $C(\omega)$ is broadened by frequent collisions and the associated spectral weight redistributed to high energies due to strong correlations. As a corollary, the featureless low frequency $C(\omega)$ is a characteristic of a bad metal with $\hbar/\tau \simeq  t \gg k_BT$  \cite{GunnarssonRMP2003,Hussey}, which gives rise to low temperature strange metal conductivity and Planckian dissipation through $P(\omega)$.\\ 

\noindent\textbf{Strange response in the cuprates, and the emergence of normal carriers}\\
With these results in mind, we can now track the evolution of metallicity in the superconducting cuprates across the phase diagram. Fig. \ref{fig:Fig2} reports the temperature evolution of the current-current correlation function $C(\omega)$ at different doping levels of Bi$_{2-x}$Pb$_{x}$Sr$_{2-y}$La$_{y}$CuO$_{4+\delta}$ (Bi2201), from an underdoped sample within the pseudogap phase (UD10K) to heavily overdoped (OD0K) beyond the edge of the superconducting dome \cite{vanHeumen22}. The corresponding \gls*{sig} is reproduced in Extended Data Figure \ref{figE:sigSM}. 

The evolution of $C(\omega)$ with both doping and temperature provides several insights that are not immediately obvious from \gls*{sig} alone. Starting with the underdoped sample UD10K, Fig. \ref{fig:Fig2}\textbf{a}, the shape of $C(\omega)$ is similar to the UD30K case (shown in Fig. \ref{fig:Fig1}\textbf{h}, \ref{fig:Fig2}\textbf{b}), but with less temperature dependence and a sharper distribution of spectral weight in the \gls*{MIR} (narrower MIR peak). Interestingly, the low temperature $C(\omega)$ for the UD10K sample shows a faster decay towards zero frequency instead of the flat response seen for the UD30K sample. This `suppression' becomes visible in the data below 250 K, which agrees with the known pseudogap temperature for this sample \cite{kondo2011disentangling}. We note that while signatures of the pseudogap in optical spectra are well established for the out-of-plane response, one has to resort to extensive modelling of \gls*{sig}  for its observation in the in-plane response \cite{Basov:RMP2011}. While more work is needed to ascertain that the suppression of $C(\omega)$ in Fig. \ref{fig:Fig2}\textbf{a} is indeed associated with the opening of the pseudogap, this would further strengthen the case for using $C(\omega)$ over \gls*{sig}.

Upon doping, the pseudogap feature disappears (Fig. \ref{fig:Fig2}B), resulting in the temperature independent extrapolated $C(\omega\to 0)$ characteristic of the strange metal. At the same time, the width of the \gls*{MIR} distribution progressively increases, in particular at elevated temperatures (Fig. \ref{fig:Fig2}B,C,D). However, the most notable feature for overdoped samples is that a second component emerges in the current correlations at low temperature. This component decreases with $\omega$, resembling the current correlations of normal carriers as shown in Fig. \ref{fig:Fig1}D,F. Its onset coincides with the doping region where a normal $T^2$ component emerges in the resistivity \cite{Ayres21,vanHeumen22}. 
\begin{figure*}[t]
    \centering
    \includegraphics[width=16cm]{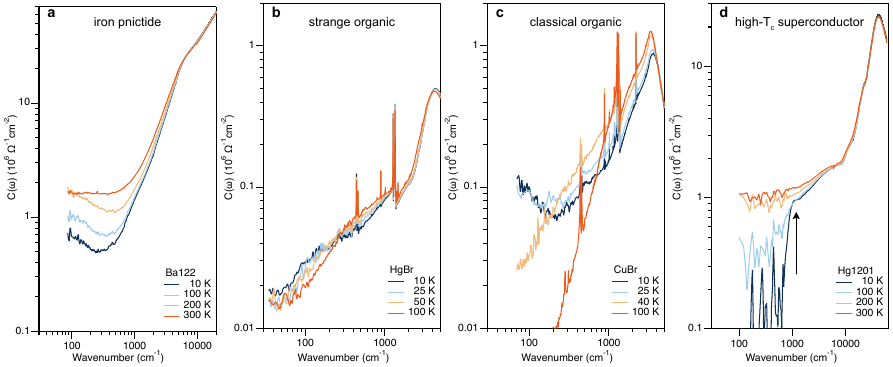}
    \caption{{\bf Current-current correlations in correlated materials} 
    \textbf{a}:~The iron-pnictide superconductor BaFe$_{1.8}$Co$_{0.2}$As$_{2}$ with $T_\text{c} = 25$~K exhibits a clear Fermi liquid response below $T\approx 150$ K \cite{tytarenko2015}. $C(\omega)$ shows a characteristic upturn at low energy, which is washed out as temperature rises. The steep rise at high wavenumber is associated with low-lying interband transitions. 
    \textbf{b}:~The organic superconductor $\kappa$-(BEDT-TTF)$_4$Hg$_{3-\delta}$Br$_{8}$ with incommensurate anions exhibits all hallmarks of a strange-metal in transport and optical behaviour \cite{Priya2027}. 
    \textbf{c}:~In contrast, $\kappa$-(BEDT-TTF)$_2$Cu[N(CN)$_2$]Br is a well-established Fermi-liquid that shows the characteristic low-temperature upturn for a coherent response (compare to panel \textbf{a}) before it turns superconducting at $T_c\approx 10$~K \cite{Priya2027}.
    \textbf{d}:~The single-layer mercury-based cuprate HgBa$_{2}$CuO$_{4+\delta}$ is a superconductor with $T_{\text{c}}\approx 97$~K \cite{heumenPRB07}. At $T=20$~K the superconducting gap with $2\Delta\approx610$ cm$^{-1}$ is clearly seen in the spectrum as well as the coherence peak indicated by an arrow, evolving from a typical strange metal spectrum at higher temperatures. The extended frequency range also includes interband transitions, which in all cases are confined at high frequencies and do not play a role in transport. 
}
    \label{fig:Fig3}
\end{figure*}

From this analysis, we can draw a rather simple picture of the evolution of metallic behaviour in the cuprates: the strange metal response is present at all doping levels and a second, classical component develops above optimal doping, closely matching what has been inferred from resistivity  \cite{Cooper09,Ayres21} and angle-resolved photoemission spectroscopy \cite{fournier2010,Chen:science2019}. 
This evolution is completely invisible in the optical conductivity itself (extended data Fig. \ref{figE:sigSM}), where the presence of $\omega \approx 0$ peaks at all doping levels has been broadly misinterpreted as a signature of long-lived carriers.\\

\noindent\textbf{Characterizing materials through $C(\omega)$}\\
The clarity gained through the analysis of $C(\omega)$ makes a compelling case to its utility. When the same lens is applied to other quantum materials, several patterns emerge as we now discuss. For instance, in the multi-orbital iron-pnictide superconductors the interaction strengths differentiate between the Fe $3d$ orbitals leading to an orbital selective Mott phase or Hund's metal \cite{de_medici2014,yu2021}. Figure \ref{fig:Fig3}\textbf{a} shows the current-current correlation function for an electron doped pnictide of the BaFe$_{2}$As$_{2}$ family at several selected temperatures. Ref. \cite{tytarenko2015} showed that the optical response had characteristic hallmarks of a \gls*{FL} with a coherence temperature $T_{0}\approx$ 1700 K and observed \gls*{FL} scaling of the optical response up to $T\approx$ 100 K. The coherent response associated with this \gls*{FL} is clearly visible in $C(\omega)$ at low frequency. As $T$ increases to room temperature, this response broadens and $C(\omega)$ flattens out into a strange metal response.

Second, Figs. \ref{fig:Fig3}\textbf{b,c} show the current correlations measured in two organic superconductors \cite{Priya2027}. 
These comprise layers of  BEDT-TTF molecules that arrange in a close-to-triangular lattice of dimers acting as effective sites. Due to the half band filling fixed by stoichiometry, the strong Coulomb interaction $U$ compared to the relatively low bandwidths $W\approx 0.5$~eV, and the absence of magnetic order, these are considered as prototypical systems to study the interaction-driven Mott transition \cite{Pustogow-NatComm}. The first compound $\kappa$-(BEDT-TTF)$_4$Hg$_{3-\delta}$Br$_{8}$ (labelled as HgBr in Fig. \ref{fig:Fig3}\textbf{b}) is very close to the idealized strange metal of Fig. \ref{fig:Fig2}\textbf{b} with almost no normal carrier response identifiable at any temperatures, which is consistently reflected in a $T$-linear resistivity \cite{Priya2027}. In contrast, the current correlations in the superconductor $\kappa$-(BEDT-TTF)$_2$Cu[N(CN)$_2$]Br ($T_\text{c}\approx 10$~K, CuBr in Fig. \ref{fig:Fig3}\textbf{c}) feature the emergence of  normal carriers at $T\le 25$K, precisely where the transport measurements indicate Fermi liquid behavior, $\rho\propto T^2$. At higher temperatures the resistivity goes through a narrow $T$-linear regime for $40<T<50$K before reaching the maximum resistivity typical of strongly correlated Mott metals at $T\simeq 100$K. Both behaviours can be read directly in the current correlations: $C(\omega)$ in CuBr at $T=40$~K is analogous to that of the strange metal HgBr, while the marked low-frequency suppression observed at $T=100$~K indicates the opening of the Mott gap. 

Fingerprints of superconductivity are equally emphasized in $C(\omega)$ as shown in Fig. \ref{fig:Fig3}\textbf{d} for the cuprate HgBa$_{2}$CuO$_{4+\delta}$ with $T_{\text{c}}\approx 97$~K \cite{heumenPRB07}. The superconducting gap size $2\Delta\approx$ 610 cm$^{-1}$ is easily estimated by the onset of absorption, while the coherence peak associated with transitions across the gap is clearly visible (indicated by the arrow). The gap is seen to close as temperature increases, but a suppression of $C(\omega)$ remains visible up to 25~K above $T_{\text{c}}$ in agreement with the fluctuation regime observed in specific heat \cite{heumenPRB07}. As the gap closes, $C(\omega)$ turns into the now familiar strange metal response.\\ 

\noindent\textbf{Discussion}\\
Our results call for a radical paradigm shift in addressing the conductivity of quantum metals. The simple transformation, equation \ref{eq:Kubo}, provides an experimental distinction between dissipation and relaxation, and at the same time it clarifies the link between strange and bad metals. The spectra of current correlations reveal the actual relaxation occurring at the microscopic level: these correlations are fast-decaying, reflecting the bad metallic character of strongly interacting systems. Combined with quantum detailed balance, this fast decay is then responsible for the strange metallicity and Planckian dissipation observed in both optical and transport measurements.

The simplicity and generality of this scenario likely explains the widespread observation of the phenomenon across different experimental platforms. At the same time, because it can clearly separate the classical and strange components, it can be used to identify materials where deviations from Fermi liquid behaviour arise through other mechanisms. For marginal Fermi liquids, in the vicinity of a true quantum critical point or when interactions are themselves disordered, measured current-current correlation functions will display structure on an energy scale smaller than or comparable to $k_{\text{B}}T$.

In the cuprates, however, this low-frequency structure is generally missing and only develops weakly at high doping levels, $p\gtrsim 0.2$ \cite{Chen:science2019,fournier2010,Ayres21}.
The strange metal response appears instead to be connected to a broad structure in $C(\omega)$ with significant frequency dependence in the \gls*{MIR}, up to energies of 1 eV. This is very reminiscent of what is found by other techniques such as electron energy loss spectroscopy and resonant inelastic x-ray scattering, that probe the same particle-hole response in a different way \cite{MitranoPNAS18,merzoni2024}. Future work should be encouraged to compare these experiments in detail and perhaps provide a unified description of all three.

From the theory standpoint it is crucial for the understanding of conduction that the full two-particle correlation function is addressed beyond estimates based on the one-particle self-energy alone. This rules out common `convolution' approximations used in dynamical mean-field theory or large-N expansions \cite{khurana1990,PatelScience23,SYK-review}, that force the low-temperature metallic behaviour to be primarily linked to the immediate vicinity of the Fermi energy. The inability of these methods to capture quantum effects is familiar in the field of localisation \cite{LeeRMP,Mahan}, a phenomenon that stems entirely from quantum interferences beyond the convolution bubble. Similarly here, accounting for processes occurring on broader momentum and energy scales, as described mathematically by vertex corrections \cite{VucicevicPRL19,Millis-PRB22,Kovacevic-PRL25,Stepanov-PRL26}, seems necessary in order to fully recover the detailed balance prefactor $P(\omega)$ \cite{FratiniNPhys26,Prelovsek-EPL01}.

Finally, our work also reinvigorates the discussion about the nature of the `glue' responsible for superconductivity in these materials \cite{anderson2007,maier2008}. 
Our data show that close to optimal doping, if there are quasi-particle carriers that couple to low-energy interactions, their contribution to the current response is negligible. 
Superconductivity is thus likely connected to the strange metallic carriers themselves, while the classical carriers appear to be antithetical to it \cite{ayres2022}. Understanding the precise nature of the high energy `elephant and mammoth' \cite{anderson2007} found here could lead to solving the high $T_{\textbf{c}}$ puzzle.\\ 

\noindent{\bf Acknowledgements}\\
\normalsize
E.v.H. thanks Blaise Gout\'eraux and Davide Valentinis for thoughtful discussions. This publication is part of the project TOPCORE with file number OCENW.GROOT.2019.048 of the research programme ENW-XL which is (partly) financed by the Dutch Research Council (NWO).
S.F. aknowledges enlightening discussions with Alberto Carta, Andrea Damascelli, Marc-Henri Julien and Luca de' Medici, and support from the French Agence Nationale de la Recherche (ANR) under grant reference number ANR-25-CE30-2817.
M.D. acknowledges support by the Deutsche Forschungsgemeinschaft (DFG) via DR 228/74-1.
\\

\noindent{\bf Competing Interests}\\
\normalsize
Authors declare no conflict of interest.\\

\noindent{\bf Data availability}\\
\normalsize
All data are available in the main text or the supplementary materials. Raw data is available upon reasonable request.\\

\noindent{\bf Code availability}\\
\normalsize
Not applicable.\\

\noindent{\bf Author contributions}\\
\normalsize
All authors contributed equally to the methodology, investigation, visualisation and writing of the paper. 

\bibliography{FlawedDrude}


\setcounter{figure}{0}
\renewcommand{\figurename}{Extended Data Figure}
\renewcommand{\thefigure}{\arabic{figure}}

\begin{figure*}[h]
    \centering
    \includegraphics[width=14cm]{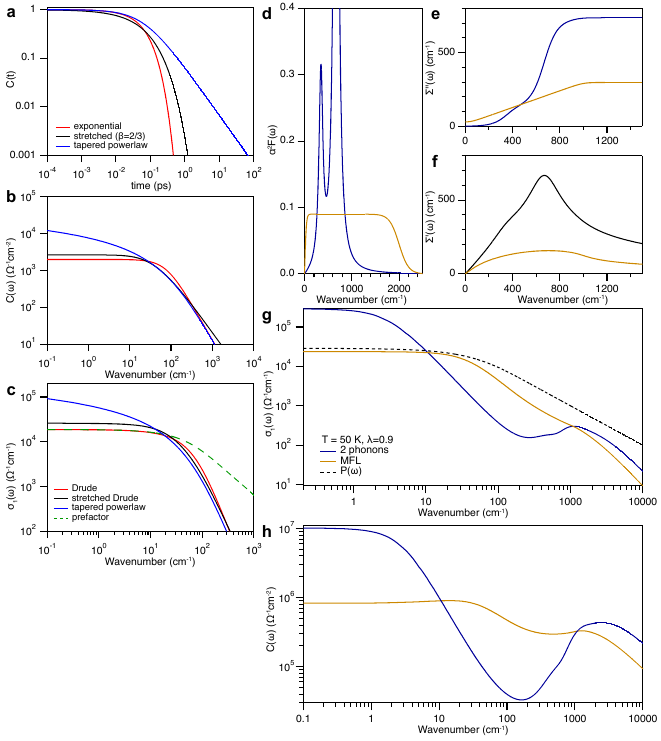}
    \caption{{\bf Time-dependent decay of current correlations and their relation to the optical conductivity}. 
    \textbf{A}:~Time dependence of C(t) for three different relaxation processes: exponential decay with $C(t)\sim\text{exp}\{-t/\tau\}$, stretched exponential decay, $C(t)\sim\text{exp}\{-(t/\tau)^\beta\} $ and a tapered powerlaw, $C(t)\sim(1+t/\tau)^{-1}$. For all three functions $\tau=67\,\text{fs}$. 
    \textbf{B}:~Corresponding frequency dependence of the correlation functions $C(\omega)$ and 
    \textbf{C}:~respective optical conductivities. For all cases the spectral weight is contained within the envelope of the prefactor and the large frequency behavior $\sim\omega^{-2}$. The different time dependence is reflected in a redistribution of spectral weight towards lower energy.\\
    The optical repsonse can also be modified through electron-boson coupling in a convolution approximation (see supplemental material section \ref{sec:eboupling}. In this case the frequency dependence is modified and spectral weight is transferred to a foot at higher energies. 
    \textbf{d}:~Electron-boson coupling function $\alpha^{2}F(\omega)$ for coupling to two phonon modes (dark blue) and a marginal Fermi liquid spectrum (MFL, brown). The integrated coupling constant $\lambda = \int\text{d}\omega\,\alpha^{2}F(\omega)/\omega$ is chosen to be the same for both spectra ($\lambda\approx$ 0.9). 
    \textbf{e,f}:~The imaginary and real parts of the self-energy $\Sigma^{\prime}(\omega)+{\rm i}\Sigma^{\prime\prime}(\omega)$ corresponding to the two cases. 
    \textbf{g}: Optical conductivity for the electron-phonon and \gls*{MFL} models at $T=$ 50 K. Both cases show a Drude-like peak with width followed by a `foot' at higher energy \cite{berthod13}. Also shown is $P(\omega)$ at this temperature (dashed black line). Despite the large coupling constant, the Fermi liquid response remains well within the envelope set by $P(\omega)$. Using the exact value of the sum rule ($\omega_{p}=$ 10\,000 cm$^{-1}$), in this example we obtain $\Gamma_{\text{sum}}/\Gamma_\text{opt} = 1.91$ for the electron-phonon case and $\Gamma_{\text{sum}}/\Gamma_\text{opt} = 1.56$ for the \gls*{MFL} spectrum. 
    \textbf{h}: Current-current correlation functions $C(\omega)$ featuring a Drude like response followed by additional structure at higher energies. The fingerprints of the phonon spectrum are clearly recognizable in $C(\omega)$, while the \gls*{MFL} foot follows the saturation of $\Sigma^{\prime\prime}(\omega)$ in panel \textbf{B}. $C(\omega)$ of the \gls*{MFL} model remains flat for low frequencies until some fraction of the temperature ($k_{\text{B}}T/\hbar\approx$ 34 cm$^{-1}$); here it slightly increases to a maximum at $\omega\approx$ 10 cm$^{-1}$. 
     The legend for all curves is shown in panel \textbf{g}. Details of the calculation are provided in section \ref{sec:eboupling} of the supplementary material.}
    \label{figE:decay}
\end{figure*}

\begin{figure*}[h]
    \centering
    \includegraphics[width=18cm]{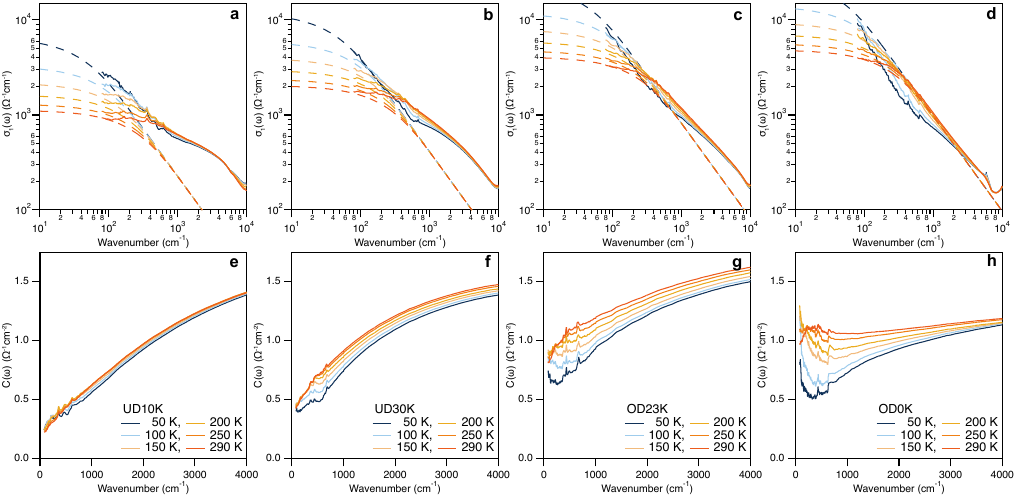}
    \caption{{\bf Optical conductivity, \gls*{sig} and current-current correlation function $C(\omega)$ across the cuprate phase diagram.} \textbf{A-D}: Optical conductivity for Bi$_{2-x}$Pb$_{x}$Sr$_{2-y}$La$_{y}$CuO$_{4+\delta}$ with different doping values corresponding to underdoped ($T_c=10$~K and $T_c=30$~K) and overdoped compounds ($T_c=23$~K and $T_c = 0$~K). The solid lines correspond to the frequency dependent conductivity $\sigma_{1}(\omega,T)$ at selected temperatures while the dashed curves represent $P(\omega)$ for the current-current correlation functions plotted in Fig. \ref{fig:Fig2} of the main text. The values of $C(0)$ used to compute the scale of the prefactor was obtained from the 50~K data for all doping levels. As doping progresses, high-frequency spectral weight from above $P(\omega)$ is transferred to low energy. \textbf{E-H}: Corresponding current-current correlation function $C(\omega)$, as plotted in Fig.~\ref{fig:Fig2} but  on linear scales.}
    \label{figE:sigSM}
\end{figure*}

\clearpage
{\Large Supplementary material}
\section{The detailed balance factor}\label{sec:detbal}
Here we briefly outline the steps that lead to Eq. \ref{eq:Kubo}. The Kubo formula is defined as:
\begin{equation}\label{condpi}
\hat{\sigma}(\omega)=\frac{\rm i}{\omega}\left[\Pi^{R}(\omega)+\frac{n_{0}e^2}{m}\right]
\end{equation}
with the retarded propagator given by,
\begin{equation}\label{eq:Pit}
\Pi^{R}(\omega) = -{\rm i}\int_{-\infty}^{\infty} {\rm d }t\, e^{-i\omega t} \Theta(t)\langle [j(t),j(0)]\rangle
\end{equation}
To proceed we need to consider the commutator $\langle [j(t),j(0)]\rangle$
Current-current correlations in real time and at finite temperature, $\langle j(t)j(0)\rangle$, are given by,
\begin{equation}
\langle j(t)j(0)\rangle=\frac{1}{Z}Tr\left[e^{-\beta H}j(t)j(0)\right]
\end{equation}
The time evolution of the current is given by $j(t)=e^{{\rm i}Ht}j(0)e^{-{\rm i}Ht}$. The Kubo-Martin-Schwinger boundary condition for thermal equilibrium states, $\langle j(0)j(t)\rangle=\langle j(t-{\rm i}\beta)j(0)\rangle$, which after Fourier transform gives:
\begin{equation}
  \langle j(0)j(\omega)\rangle=e^{-\hbar\omega/k_{b}T}\langle j(\omega)j(0)\rangle  
\end{equation}
This is valid so long as $\langle j(z)j(0)\rangle$ is analytic for $0<z<\beta$. We write,
\begin{equation}\label{eq:Pit2}
\Pi^{R}(\omega) = -{\rm i}\int_{-\infty}^{\infty}e^{-{\rm i}\omega t} \Theta(t)[\langle j(t)j(0)\rangle-\langle j(t-i\beta)j(0)\rangle]
\end{equation}
Since the Fourier transform of a product is equivalent to the convolution of the two functions in the reciprocal space, Eq. \ref{eq:Pit2} becomes after Fourier transformation,
\begin{eqnarray}
\Pi^R(\omega) &=& -{\rm i} \left( \frac{\rm i}{\omega + i0^+} * [\langle j(\omega)j(0)\rangle \right.\nonumber\\
&&\left.-e^{-\hbar\omega/k_{b}T}\langle j(\omega)j(0)\rangle]\right)\nonumber\\
&=& \frac{1}{\pi} \int_{-\infty}^{\infty}  {\rm d}\Omega \, \frac{[1 -e^{-\hbar\omega/k_{b}T}]\langle j(\omega)j(0)\rangle}{\omega - \Omega + {\rm i}0^+}
\end{eqnarray}
where we have used the Fourier transform of the step function,
\begin{equation}
\int_{-\infty}^{\infty} {\rm d}t\, \, e^{{\rm i}\omega t} \theta(t) = \frac{\rm i}{\omega + {\rm i}0^+} = {\rm i} \mathcal{P}\left(\frac{1}{\omega}\right) + \pi \delta(\omega)  
\end{equation}
Using the right hand side of this expression, we finally find for the retarded propgator,
\begin{eqnarray}\label{eq:Piw}
\Pi^R(\omega) &= \frac{1}{\pi} \mathcal{P} \int_{-\infty}^{\infty}  {\rm d}\Omega \, \frac{(1-e^{-\omega/T})\langle j(\omega)j(0)\rangle}{\omega - \Omega}\nonumber\\ 
&- {\rm i}(1-e^{-\omega/T})\langle j(\omega)j(0)\rangle
\end{eqnarray}
Combining this with Eq. \ref{condpi} and taking the real part leads to Eq. \ref{eq:Kubo} of the main paper.\\ 

\section{Optical conductivity from coupling to bosons.}\label{sec:eboupling}
In extended Figure \ref{figE:decay} we compute the optical conductivity and related current-current correlation function. Here we provide some details of these calculations. The calculation starts by defining a spectrum of bosons that couple to the electrons and give rise to a self energy using the lowest-order rainbow diagram in the limit of an infinite bandwidth
\begin{eqnarray}
\Sigma \left( {\omega ,T} \right) &=& \int d\varepsilon \int d\omega ^{\prime}B(\omega ^{\prime})\left[\frac{{n_\text{B} (\omega^{\prime}) + n_\text{F} (\varepsilon )}}{{\omega  - \varepsilon  + \omega ^{\prime} + \text{i}\delta }}\right.\nonumber\\
&&\left. + \frac{{n_\text{B} (\omega^{\prime}) + 1 - n_\text{F} (\varepsilon )}}{{\omega  - \varepsilon  - \omega^{\prime} - \text{i}\delta }}\right], 
\end{eqnarray}
where $n_\text{B}(x)$ and $n_\text{F}(x)$ are the Bose-Einstein and Fermi-Dirac
distribution functions, respectively, and $B(\omega)$ is the boson spectrum. The integral over $\varepsilon$ can be
performed analytically to obtain
\begin{equation}\label{SE2}
\Sigma \left( {\omega ,T} \right) = \int d\omega^{\prime} B(\omega^{\prime})L\left(\omega,\omega^{\prime},T\right), 
\end{equation}
where 
\begin{eqnarray}\label{Lfunc}
L(\omega ,\omega^{\prime} ,T) &=& - \text{i}\pi \coth \left( {\frac{\omega
}{{2T}}} \right) \nonumber\\
&&+ \Psi \left( {\frac{1}{2} + \text{i}\frac{{\omega  -
\omega^{\prime} }}{{2\pi k_\text{B}T}}} \right) - \Psi \left( {\frac{1}{2} -
\text{i}\frac{{\omega + \omega^{\prime} }}{{2\pi k_\text{B}T}}} \right)\nonumber
\end{eqnarray}
and $\Psi(x)$ is the digamma function. These expressions are then evaluated numerically to obtain the self-energy. 

Once the self-energy is computed, we compute the optical conductivity using the Kubo integral,
\begin{equation}
\sigma_{1}(\omega)=\frac{\omega_{p}^{2}}{4\pi {\rm i}\omega}{}\int_{0}^{\infty}\text{d}\varepsilon\frac{n_\text{F} (\varepsilon )-n_\text{F} (\varepsilon + \omega)}{\omega+\Sigma^{*}(\varepsilon)-\Sigma(\varepsilon+\omega)}
\end{equation}

For the phonon spectrum we use two Lorentzian lineshapes to define the phonon modes, while we use the following for the \gls*{MFL} spectrum,
\begin{equation}\label{eq:BMFL}
B(\omega) = g\frac{\tanh\left( \frac{\omega}{2k_\text{B}T} \right)}{1 + \exp\left( \frac{\omega - \omega_{c}}{\Delta} \right)}. 
\end{equation}
Here, $g$ is a coupling constant and the denominator is a cut-off function that smoothly terminates the boson spectrum around the energy $\omega_\text{c}$ with width $\Delta$. 

\end{document}